\documentclass[12pt]{article}
\usepackage{epsfig}
\usepackage{amsmath}

    \def\z{\noindent}  
 
    \def\sqr#1#2{{\vcenter{\vbox{\hrule height .#2pt
                             \hbox{\vrule width .#2pt height#1pt \kern#1pt
                                   \vrule width .#2pt}
                             \hrule height .#2pt}}}}

    \def\NN{\bf{N}}
    
    \def\RR{\bf{R}}
    \def\ZZ{\bf{Z}}

\begin{document}
{ \title{Exact Results for the Ionization of a Model Quantum
System}

\author{O. Costin, J. L. Lebowitz\footnote{Also Department of Physics}, and
A. Rokhlenko \\ Department of Mathematics\\ Rutgers University
\\ Piscataway, NJ 08854-8019}

\maketitle
\begin{abstract}
 We prove that a model atom having one bound state will be fully ionized
  by a time periodic potential of arbitrary strength $r$ and frequency
  $\omega$.  Starting with the system in the bound state, the survival
  probability is for small $r$ given by $e^{-\Gamma t}$ for times of
  order $\Gamma^{-1}$ $\sim r^{-2n}$, where $n$ is the minimum number of
  ``photons'' required for ionization (with large modifications at
  resonances).  For late times the decay is like $t^{-3}$ with the power
  law modulated by oscillations. As $r $ increases the time over which
  there is exponential decay becomes shorter and the power law behavior
  starts earlier.  Results are for a parametrically excited one
  dimensional system with zero range potential but comparison with
  analytical works and with experiments indicate that many features are
  general.

\smallskip

\z PACS: 32.80 Rm, 03.65 Bz, 32.80 Wr.
\end{abstract}

\centerline{*******}}

\section{Introduction}

The solution of the Schr{\"o}dinger equation with a time dependent
potential leading to transitions between bound and free states of a
quantum system is clearly of great theoretical and practical interest.
Fermi's golden rule (based on a deep physical interpretation of first
order perturbation theory) gives the decay exponent of the survival
probability for a system in a bound state subjected to a weak external
oscillating potential, with frequency $\omega > \omega_0 = -u_b/\hbar$,
$u_b$ the energy of the bound state \cite{[1]}.  The approaches used to
go beyond the golden rule include higher order perturbation theory,
semi-classical phase-space analysis, Floquet theory, complex dilation,
exact results for small fields, and numerical integration of the time
dependent Schr{\" o}dinger equation \cite{[1]}-\cite{[14]}.  These works
have yielded both theoretical understanding and good agreement with
dissociation experiments in strong laser fields.  In particular they
have been very successful in elucidating much of the rich structure
found in the experiments on the multiphoton ionization of Rydberg atoms
by microwave fields \cite{[2]}--\cite{[6]}.  Explicit results for
realistic systems require, of course, the use of some approximations
whose reliability is not easy to establish a priori.

It would clearly  be desirable to have examples for which one could
compute the time evolution of an initially bound state and thus the
ionization probability for all values of the frequency and strength of the
oscillating potential to as high accuracy as desired without any {\it
uncontrolled} approximations.  This is the motivation for the present work
which describes new exact results relating to ionization of a
very simple model atom by an oscillating field (potential) of arbitrary
strength and frequency.  While our results hold for arbitrary strength
perturbations, the predictions are particularly explicit and sharp in
the case where the strength of the oscillating field is small relative
to the binding potential---a situation commonly encountered in practice.
Going beyond perturbation theory we rigorously prove the existence of a
well defined exponential decay regime which is followed, for late times
when the survival probability is already very low, by a power law decay.
This is true no matter how small the frequency. The times required for
ionization are however very dependent on the perturbing frequency.  For
a harmonic perturbation with frequency $\omega$ the logarithm of the
ionization time grows like $r^{-2n}$, where $r$ is the normalized
strength of the perturbation and $n$ is the number of ``photons''
required for ionization.  This is consistent with conclusions drawn from
perturbation theory and other methods (the approach in \cite{[7a]} being
the closest to ours), but is, as far as we know, the first exact result
in this direction.  We also obtain, via controlled schemes, such as
continued fractions and convergent series expansions, results for strong
perturbing potentials.

Quite surprisingly our results reproduce many features of the
experimental curves for the multiphoton ionization of excited hydrogen
atoms by a microwave field \cite{[3]}.  These features include both the
general dependence of the ionization probabilities on field strength as
well as the increase in the life time of the bound state when
$-n\hbar\omega$, $n$ integer, is very close to the binding energy. Such
``resonance stabilization'' is a striking feature of the Rydberg level
ionization curves [3].  These successes and comparisons with analytical
results \cite{[1]}-\cite{[10]} suggest that the simple model we shall now
describe contains many of the essential ingredients of the ionization
process in real systems.

\subsection{Description of the model}

 The unperturbed Hamiltonian in our model is,

\begin{equation}
  \label{eq:(1)}
  H_0=-{\frac{\hbar^2}{2m}}{\frac{d^2}{dy^2}}-g\delta (y),\ g>0,\ \ -\infty<y<\infty.
\end{equation}
 $H_0$ has a single bound state $ u_b(y)=\sqrt{p_0}e^{-p_0|y|},\
p_0=\frac{m}{\hbar^2}g$ with energy $-\hbar \omega_0=-\hbar^2p_0^2/2m$ and a
continuous uniform spectrum on the positive real line, with  generalized
eigenfunctions 
$$u(k,y)=\frac{1}{\sqrt{2\pi}}\left
(e^{iky}-\frac{p_0}{p_0+i|k|}e^{i|ky|} \right ), \ \ -\infty<k<\infty$$

\z and energies $\hbar^2k^2/2m$.  

Beginning at some initial time, say  $t=0$, we apply a parametric perturbing
potential  $ -  g\eta(t) \delta(y)$, i.e.\  we  change the parameter $g$ in
$H_0$  to  $g(1 + \eta(t))$  and  solve the time  dependent Schr{\"o}dinger
equation for $\psi(y,t)$,

\begin{equation}
  \label{eq:(2)}
 \psi (y,t)=\theta (t)u_b(y)e^{i \omega_0 t}+\int_{-\infty}^
{\infty}\Theta (k,t)u(k,y)e^{-i\frac{\hbar k^2}{2m}t}dk \ \ (t\geq 0)
\end{equation}
with initial values $\theta (0)=1,\ \Theta (k,0)=0$.  This gives the
survival probability $|\theta(t)|^2$, as well as the fraction of ejected
electrons $|\Theta(k,t)|^2 dk$ with (quasi-) momentum in the interval
$dk$.

This problem can be reduced
to the solution of an integral equation \cite{[15]}.   Setting 

\begin{eqnarray}
  \label{eq:(3)}
  &\displaystyle \theta (t)=1+2i\int_0^t Y(s) ds \\
  &\displaystyle \Theta(k,t)= \frac{2|k|}{\sqrt{2\pi} (1-i|k|)}\int_0^t Y(s)
e^{i(1+k^2)s} ds 
\end{eqnarray}

\z  $Y(t)$ satisfies the integral equation

\begin{equation}
  \label{eq:(5)}
  Y(t)=\eta(t)\left
\{1+\int_0^t [2i+M(t-t')]Y(t') dt'\right \}
\end{equation}

\z where we have set $p_0=\omega_0=\hbar=2m=\frac{g}{2}=1$ and
$$ M(s)=\frac{2i}{\pi}\int_0^\infty \frac{u^2e^{-is(1+u^2)}}{1+u^2}du=
\frac{1}{2}\sqrt{\frac{i}{\pi}}\int_s^\infty\frac{e^{-iu}}{u^{3/2}} du.
$$ 

\section{Results}

Our first exact result is the following: {\em When $\eta(t)$ is a
trigonometric polynomial, 

\begin{equation}
  \label{eq:(7)}
 \eta(t)=\sum_{j=1}^{n}[A_j\sin(j\omega
  t)+B_j\cos(j\omega t)],
\end{equation}
\z  the survival probability $|\theta(t)|^2$ tends to
zero as $t\rightarrow\infty$, for all $\omega>0$}.  
That is there will be
full ionization for arbitrary strength and frequency of the oscillating
field.   

 Since the main features of the argument are already present in the
simplest case $\eta=r\sin(\omega t)$ we now specialize to this case.  The
asymptotics of $Y(t)$ are obtained from its Laplace transform $y(p) =
\int^\infty_0 e^{-pt} Y(t) dt$, which satisfies the functional equation
(cf. (\ref{eq:(5)}))

\begin{equation}
  \label{eq:(8)}
y(p) = \frac{ir}{2}\left\{{\frac{y(p+i\omega )}{ \sqrt{1-ip+\omega} - 1}} -
{\frac{y(p-i\omega)}{\sqrt{1-ip-\omega}-1}}\right\}+
{\frac{r\omega}{
\omega^2+p^2}}
\end{equation}
 \z with the boundary condition $y(p)\rightarrow 0$ as
$\Im(p)\rightarrow \pm\infty$ (the relevant branch of the square root is
$(1-ip-\omega)^{1/2} = -i(\omega-1+ip)^{1/2}$ for $\omega > 1$). We show
that the solution of (\ref{eq:(8)}) with the given boundary conditions is
unique and analytic for $\Re(p)>0$, and its only singularities on the
imaginary axis are square-root branch points.  This in turn implies that
$|Y(t)|$ does indeed decay in an integrable way.  

The ideas of the proof carry through directly to the more general
periodic potential (\ref{eq:(7)}) and we have obtained analogous results
for a two delta functions reference potential \cite{[16]}.

Full ionization is in fact expected (for entropic reasons) to hold
generically, but has, as far as we know, only been proven before for
small amplitude of the oscillating potential with $\omega>1$
(\cite{[9]}, \cite{[10]}), or for sufficiently random perturbations
(\cite{[9]}).

The detailed behavior of the system as a function of $t$, $\omega$, and
$r$ is obtained from a precise study of the singularities of $y(p)$ in
the whole complex $p$-plane. Here we discuss them {\bf for small $r$};
below, the symbol $\varepsilon $ describes error bounds of order
$O(r^{2-\delta}), \frac{1}{2}<\delta<1$. At $p=\{i n\omega-i:
n\in\ZZ\}$, $y$ has square root branch points and $y$ is analytic in the
right half plane and also in an open neighborhood ${\cal{N}}$ of the
imaginary axis with cuts through the branch points. As
$|\Im(p)|\rightarrow\infty$ in ${\cal{N}}$ we have $|y(p)|=O(r \omega
|p|^{-2})$.  If $|\omega-\frac{1}{n}|> C_n\varepsilon ,\,n\in\ZZ^+$, for
some constants $C_n$, then for small $ r $ the function $y$ has a unique
pole $p_m$ in each of the strips $ -m\omega>\Im(p)+1\pm \varepsilon
>-m\omega-\omega,\,\ m\in\ZZ$.  $\Re(p_m)$ is strictly independent of
$m$ and gives the exponential decay of $\theta$.

\begin{figure}
\epsfig{file=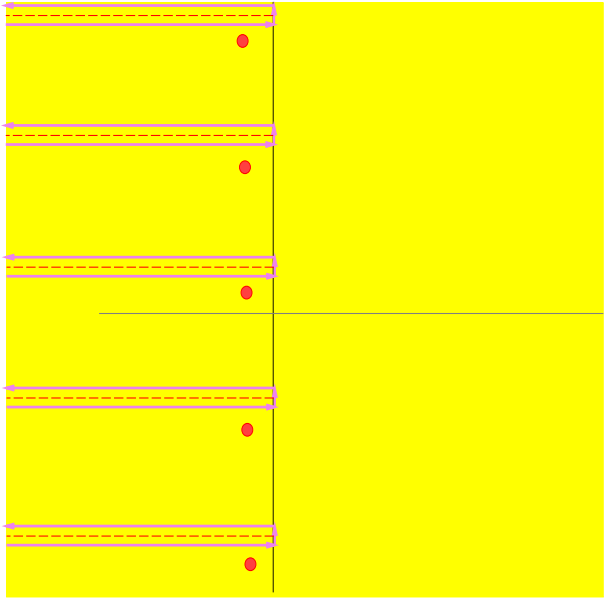, width=8cm,height=8cm}
\caption{ Singularities of $y$ and relevant inverse Laplace contours.}
\end{figure}

 The analytic structure of $y$ is indicated in Figure
1 where the dotted lines represent (the square root) branch cuts and the
dark circles are simple poles. The function $Y$ is the inverse Laplace
transform of $y$

\begin{equation}\label{6}
Y(t)=\frac{1 }{ {2 \pi i}} ~\int_{{\mathcal{C}}} 
e^{p t }y(p) ~dp 
\end{equation}

\z where the contour of integration ${\mathcal{C}}$ can be initially taken to
be the imaginary axis $i\RR$, since $y$ is continuous there and decays
like $p^{-2}$ for large $p$.

We then show that $\mathcal{C}$ can be pushed through the poles,
collecting the appropriate residues, and along the branch cuts as shown
in Figure 1. The residue at the pole $p_m$ is proportional to
$e^{(p_0+im\omega) t}$ while the (rapidly convergent) integral along the
$m-$th 
branch cut is (as seen by standard Laplace integral techniques), a
function whose large $t$ behavior is $K_m e^{i m\omega t} t^{-3/2}$
(the $-3/2$ power comes from the fact that $y$ has square root branch
points; $K_m$ is some constant). 

A detailed analysis along these lines 
yields (\cite{[16]}, \cite{[16a]}), 

\begin{eqnarray}
  \label{eq:intform}
 \theta(t)=
e^{-\gamma(r;\omega) t}F_\omega(t)+\sum_{m=-\infty}^\infty
e^{(mi\omega-i)t}h_m(t)&
\end{eqnarray}

\z where $F_\omega$ is periodic of period $2\pi\omega^{-1}$ and
$h_m(t)\sim \sum_{j=0}^{\infty}c_{m,j}t^{-3/2-j}$ for large $t$ in more
than a half plane centered on the positive real half-line. {\em Not too
close to resonances}, i.e.\ when $|\omega-n^{-1}|>\varepsilon$, for all
integer $n$, $|F_{\omega}(t)|=1+O(r^2)$ and its Fourier coefficients
decay faster than $r^{|2m|} |m|^{-|m|/2}$. Also, the sum in
(\ref{eq:intform}) does not exceed $O(r^2 t^{-3/2})$ for large $t$, and
the $h_m$ decrease with $m$ faster than $r^{|m|}$.

\begin{figure}
\epsfig{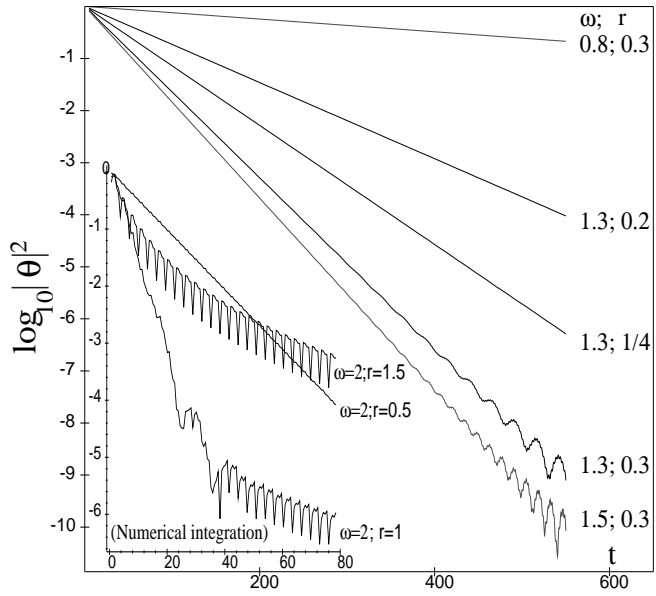}
\caption{Plot of $\log_{10}|\theta(t)|^2$ vs. time in units of
$\omega_0^{-1}$ for several values of $\omega$ and $r$. The main graph
was calculated from (8) and the inset used numerical integration of (5).}
\end{figure}

By (\ref{eq:intform}), for times of order $1/\Gamma$ where
$\Gamma=2\Re(\gamma)$, the survival probability for $\omega$ not close
to a resonance decays as $\exp(-\Gamma t)$. This is illustrated in
Figure 2 where it is seen that for $r^{<}\hskip-1.5ex_{\sim} 0.5$ the
exponential decay holds up to times at which the survival probability is
extremely small.  Note also the slow decay for $\omega = .8$, when
ionization requires the absorption of two photons. For even larger $r$
one can note in the figure oscillatory behavior. This is expected from
equation (\ref{eq:intform}).

When $r$ is larger (inset in Fig. 2) the ripples of $|F_{\omega}(t)|$
are visible and the polynomial-oscillatory behavior starts sooner. Since
the amplitude of the late asymptotic terms is $O(r^2)$ for small $r$,
increased $r$ can yield a higher late time survival probability. This
phenomenon, sometimes referred to as ``adiabatic stabilization''
\cite{[12a]},  \cite {[17]}, can be associated with the
perturbation-induced probability of back-transitions to the well.

Using continued fractions $\Gamma$ can be calculated convergently for
any $\omega$ and $r$. For small $r$ we have

\begin{equation}
\Gamma\sim\left\{\begin{array}{lllllll} \displaystyle
      \sqrt{\omega-1}\frac{ r ^2}{\omega}; &\mbox{\rm if }{\omega}\in (1+\varepsilon,\infty)
      \\ &\\
  \displaystyle 
\frac{ \sqrt{2\omega -1}}{(1-\sqrt{1-\omega})^2}\frac{ r
  ^4}{8\omega};&\mbox{\rm if }{\omega} \in (\frac{1}{2}+\varepsilon,1-\varepsilon)
\\
\ldots &\ldots \label{devilstaircase}
\\ \displaystyle\frac{2^{-2n+2}\sqrt{n\omega-1}}{\displaystyle\prod_{m<
n}(1-\sqrt{1-m\omega})^2}\frac{ r ^{2n}}{n\omega}; &\mbox{\rm if }{\omega}\in
(\frac{1}{n}+\varepsilon,\frac{1}{n-1}-\varepsilon)
\end{array}\right.\end{equation}

Eq. (\ref{devilstaircase}) agrees with
results of perturbation theory.  Thus, for $\omega > 1$, $\Gamma$ is
given by Fermi's golden rule \cite{[1]} since the transition matrix
element between the bound state with energy $-1$ and the continuum state
with energy $k^2$ is

\begin{equation}\label{e10}
\big|<u_b(y)|\delta(y)u(k,y)>\big|^2=\frac{1}{2\pi}\frac{k^2}{1+k^2}
\end{equation}
while the density of states is ${\frac{2\pi}{k}}$.  

The behavior of $\Gamma$ is different at the resonances
$\omega^{-1}\in\NN$. For instance, whereas if $\omega$ is not close to
$1$, the scaling of $\Gamma$ implied by (\ref{devilstaircase}) is $r^2$ when
$\omega>1$ and $r^4$ when $\frac{1}{2}<\omega<1$, when
$\omega-1=r^2/\sqrt{2}$ we find 

$$\Gamma\sim \Big(\frac{2^{1/4}}{8}-\frac{2^{3/4}}{16}\Big)r^3$$

In Figure 3 we plot the behavior of $\Gamma^{-1}$, as a function of
$\omega$, for a small value of $r$. The curve is made up of smooth
(roughly self-similar) pieces for $\omega$ in the intervals $(n^{-1},
(n-1)^{-1})$ corresponding to ionization by $n$ photons.
\begin{figure}
\epsfig{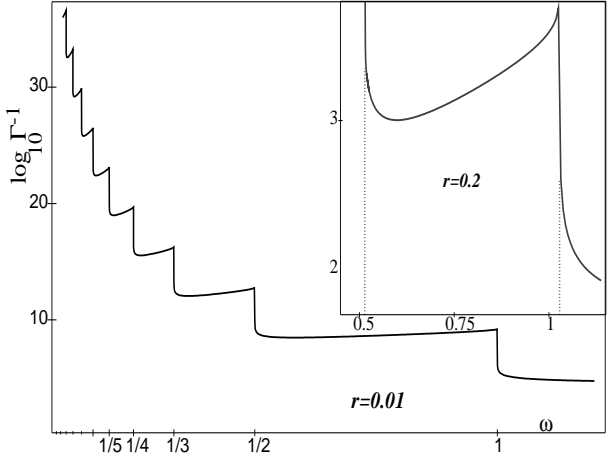}

\caption{$\log_{10}\Gamma^{-1}$ as a function of $\omega/\omega_0$ at
$r=0.01$. At $r=0.2$ (inset) shift of the resonance is visible. In the
inset the continued fraction was used, while for the main figure
the data was extrapolated from (\ref{devilstaircase}).}

\end{figure}
At resonances (for small $r$ these occur for $\omega^{-1}$ close to an
integer), the coefficient of $r^{2n}$, the leading term in $\Gamma$,
goes to zero.  This yields an enhanced stability of the bound state
against ionization by perturbations with such frequencies. The origin of
this behavior is, in our $d=1$ model, the vanishing of the matrix
element in (\ref{e10}) at $k=0$. This behavior should hold quite
generally in $d = 3$ where there is a factor $k$ in $\Gamma$ coming from
the energy density of states near $k=0$.  As $r$ increases these
resonances shift in the direction of increased frequency.  For small $r$
and $\omega=1$ the shift in the position of the resonance, sometimes
called the dynamic Stark effect [1], is about $\frac{r^2}{\sqrt{2}}$.

In Figure 4 we plot the strength of the perturbation $r$, required to
make $|\theta(t)|^2 = {\frac{1}{2}}$ for a time of about 700 oscillations
of the perturbing field, i.e.\ time is measured in units of
$\omega^{-1}$, as a function of $\omega$.  Also included in this figure
are experimental results taken from Table 1 in \cite{[3]}, see also
Figures 13 and 18 there for the ionization of a hydrogen atom by a
microwave field for approximately the same number of oscillations.  In
these still ongoing beautiful series of experiments,
\cite{[3]}--\cite{[5]}, the atom is initially in an excited state with
principal quantum number $n_0$ ranging from 32 to 90.  The ``natural
frequency'' $\omega_0$ is there taken to be that of an average
transition from $n_0$ to $n_0 \pm1$, so $\omega_0 \sim n^{-3}_0$.  The
strength of the microwave field $F$ is then normalized to the strength
of the nuclear field in the initial state, which scales like $n^{-4}_0$.
The plot there is thus of $n^4_0 F$ vs. $n^3_0 \omega$.  To compare the
results of our model with the experimental ones we had to relate $r$ to
$n_0^4 F$, and given the difference between the hydrogen atom perturbed
by a polarized electric field $V_1=xF\sin(\omega t)$, and our model,
this is clearly not something that can be done in a unique way.  We
therefore simply tried to find a correspondence between $n_0^4F$ and $r$
which would give the best visual fit. Somewhat to our surprise these
fits for different values of $\omega/\omega_0$ all turned out to have
values of $r$ close to $3n_0^4F$.

\begin{figure}
\epsfig{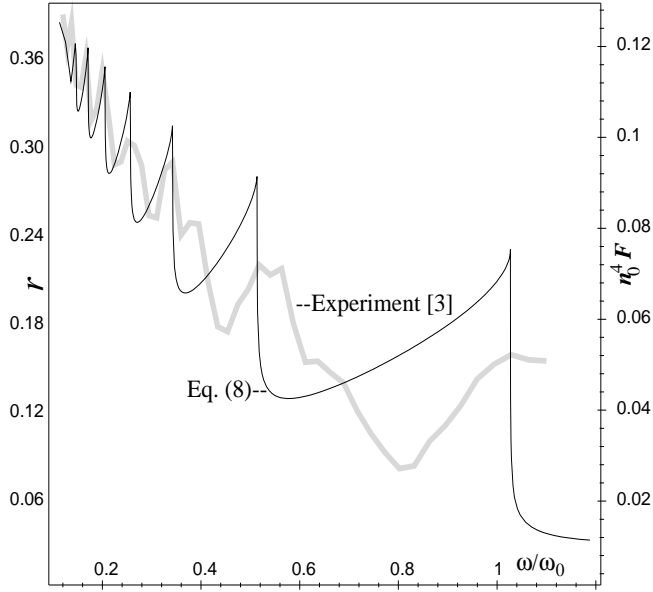}
\caption{Threshold amplitudes for $50\%$ ionization
vs. $\omega/\omega_0$, calculated from eq. (\ref{devilstaircase}), and in the experiment [3]. In
the calculation, the dynamic Stark effect was approximated using an
averaged $r$ over the range.}

\end{figure}

A correspondence of the same order of magnitude is obtained by
comparing the perturbation-induced shifts of bound state energies in our
model and in hydrogen.  We note that the maximal values of $r$ in Figure\ 4
are still within the regime where only a few terms in (\ref{eq:intform})
are sufficient.

In Figure~5 we plot $|\theta(t)|^2$ vs. $r$ for a fixed $t$ and two
different values of $\omega$. These frequencies are chosen to correspond to
the values of $\omega/\omega_0$ in the experimental curves,  Figure 1 in
\cite{[5]} and Figure 1b in \cite{[3]}. The agreement is very good for
$\omega/\omega_0\approx .1116$ and reasonable for the larger ratio.  Our
model essentially predicts that when the fields are not too strong, the
experimental survival curves for a fixed $n_0^3 \omega$ (away from the
resonances) should behave essentially like $\displaystyle\exp\left(-C[n^4_0
F]^{\scriptstyle\frac{\scriptstyle 2}{\scriptstyle
n_0^{3}\omega}}\,\omega\,t\right)$ with $C$ depending on $n_0^3\omega$ but,
to first approximation, independent of $n_0^4 F$.

\begin{figure}\label{fig5}
\epsfig{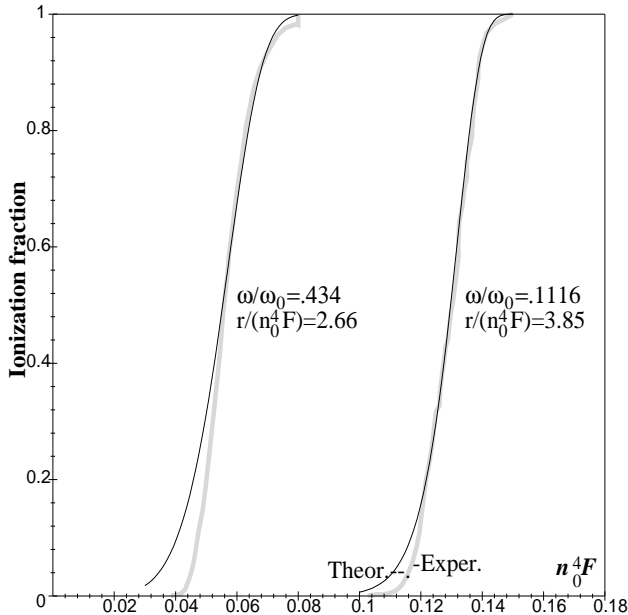}
\caption{Ionization fraction at fixed $t$ (corresponding to 300
oscillations) as a function of amplitude.}
\end{figure}

\section{Concluding remarks}

Given the simplicity of our model, the similarity (using a minimal
numbers of adjustable parameters) with the experiments on hydrogen atoms
was surprising to us.  As already noted these experimental results and in
particular the resonances can be understood quite well, including some
details, by doing calculations on the full hydrogen atom or a one
dimensional version of it \cite{[2]}-\cite{[6]}.  Still, it is
interesting to see that similar structures arise also in very simple
models.  It suggests that various features of the ionization process
have a certain universal character. To really pin down the reason for
this universality will require much further work.

We note that for $\omega>\omega_0$, in the limit of small amplitudes
$r$, a predominantly exponential decay of the survival probability
followed by a power-law decay was proved in \cite{[10]} for three
dimensional models with quite general local binding potentials having
one bound state, perturbed by $r\cos(\omega t)V(y)$, where $V$ is a local
potential. Our results for general $\omega$ and $r$ can be thought of as
coming from a rigorous Borel summation of the formal (exponential)
asymptotic expansion of $Y$ for $t\rightarrow\infty$. These methods can
be extended to other systems \cite{[15]} including, we hope, realistic
ones.

\z {\bf Acknowledgments}.  We thank A. Soffer, M. Weinstein and P. M. Koch for
valuable discussions and for providing us with their papers. We also
thank R. Barker, S. Guerin and H. Jauslin for introducing us to the
subject.  Work of O. C. was supported by NSF Grant 9704968, that of J.
L. L. and A. R. by AFOSR Grant F49620-98-1-0207.

\smallskip

\z costin@math.rutgers.edu\hfill\break
 lebowitz@sakharov.rutgers.edu\hfill\break
rokhlenk@math.rutgers.edu.

\end{document}